%Paper: hep-th/9205002
%From: kawai@theory.kek.jp
%Date: Fri, 01 May 92 22:14:22 JST

%%%%%%%%%%%%%%%%%%%%%%%%%%%%%%%%%%%%%%%%%%%%%%%%%%%%%%%%%%%%%%%%%%%%%%%
%%
%
%             Fusion Rules for the $q$-deformed Vertex Operators
%
%                      of $U_q(\widehat{sl(2)})$
%
%                            H. Awata
%
%                            Y. Yamada
%
%                         Written by KEK Phyzzx
%
%%%%%%%%%%%%%%%%%%%%%%%%%%%%  Abstract  %%%%%%%%%%%%%%%%%%%%%%%%%%%%%%%
%%
%     We analyze the properties of the $q$-vertex operators of
%    $U_q(\widehat{sl(2)})$ introduced by Frenkel and Reshetikhin.
%    As the condition for the null vector decoupling,
%    we derive the existence condition of the $q$-vertex operators
%    ( the fusion rules ).
%
%%%%%%%%%%%%%%%%%%%%%%%%%%%%%%%%%%%%%%%%%%%%%%%%%%%%%%%%%%%%%%%%%%%%%%%%
%%%   This is PHYZZX macro package.   % % % % % % % % % % % % % % % % %%
%%%  This version of PHYZZX should be used with Version 1.0 of TEX  % %%
%%%   Do not "\input phyzzx" unless you preload or "\input" PLAIN.  % %%
%%%   To preload both PLAIN and PHYZZX, begin your file with	% % % %%
%%%  a line "%macropackage=phyzzx" instead of "\input phyzzx".  % % % %%
%   Modified by K. Hikasa: Jan. 9, 1992 Version                        %
%%%%%%%%%%%%%%%%%%%%%%%%%%%%%%%%%%%%%%%%%%%%%%%%%%%%%%%%%%%%%%%%%%%%%%%%
%%%%%%%  Created by Vadim Kaplunovsky in June 1984.   %%%%%%%%%%%%%%%%%%
%%%%%%%%%%%%  Latest update/debug: September 24, 1984.	  %%%%%%%%%%%%%%
%%%%%%%%%%%%%%%%%%%%%%%%%%%%%%%%%%%%%%%%%%%%%%%%%%%%%%%%%%%%%%%%%%%%%%%%
\catcode`@=11 % This allows us to modify PLAIN macros.
%%%%%%%%%%%%%%%%%%%%%%%%%%%%%%%%%%%%%%%%%%%%%%%%%%%%%%%%%%%%%%%%%%%%%%%%
%   I begin with fonts.

\font\fourteenrm=cmr10 scaled\magstep2
\font\twelverm=cmr10 scaled\magstep1
\font\ninerm=cmr9	     \font\sixrm=cmr6

\font\fourteenbf=cmbx10 scaled\magstep2
\font\twelvebf=cmbx10 scaled\magstep1
\font\ninebf=cmbx9	      \font\sixbf=cmbx6
\font\seventeeni=cmmi10 scaled\magstep3	    \skewchar\seventeeni='177
\font\fourteeni=cmmi10 scaled\magstep2	    \skewchar\fourteeni='177
\font\twelvei=cmmi10 scaled\magstep1	    \skewchar\twelvei='177
\font\ninei=cmmi9			    \skewchar\ninei='177
\font\sixi=cmmi6			    \skewchar\sixi='177
\font\seventeensy=cmsy10 scaled\magstep3    \skewchar\seventeensy='60
\font\fourteensy=cmsy10 scaled\magstep2	    \skewchar\fourteensy='60
\font\twelvesy=cmsy10 scaled\magstep1	    \skewchar\twelvesy='60
\font\ninesy=cmsy9			    \skewchar\ninesy='60
\font\sixsy=cmsy6			    \skewchar\sixsy='60

\font\fourteenex=cmex10 scaled\magstep2
\font\twelveex=cmex10 scaled\magstep1

\font\fourteensl=cmsl10 scaled\magstep2
\font\twelvesl=cmsl10 scaled\magstep1
\font\ninesl=cmsl9

\font\fourteenit=cmti10 scaled\magstep2
\font\twelveit=cmti10 scaled\magstep1
\font\twelvett=cmtt10 scaled\magstep1
\font\twelvecp=cmcsc10 scaled\magstep1
\font\tencp=cmcsc10
\newfam\cpfam
	% quick fix for a missing font
\newcount\f@ntkey	     \f@ntkey=0
\def\samef@nt{\relax \ifcase\f@ntkey \rm \or\oldstyle \or\or
	 \or\it \or\sl \or\bf \or\tt \or\caps \fi }
\def\fourteenpoint{\relax
    \textfont0=\fourteenrm	    \scriptfont0=\tenrm
    \scriptscriptfont0=\sevenrm
     \def\rm{\fam0 \fourteenrm \f@ntkey=0 }\relax
    \textfont1=\fourteeni	    \scriptfont1=\teni
    \scriptscriptfont1=\seveni
     \def\oldstyle{\fam1 \fourteeni\f@ntkey=1 }\relax
    \textfont2=\fourteensy	    \scriptfont2=\tensy
    \scriptscriptfont2=\sevensy
    \textfont3=\fourteenex     \scriptfont3=\fourteenex
    \scriptscriptfont3=\fourteenex
    \def\it{\fam\itfam \fourteenit\f@ntkey=4 }\textfont\itfam=\fourteenit
    \def\sl{\fam\slfam \fourteensl\f@ntkey=5 }\textfont\slfam=\fourteensl
    \scriptfont\slfam=\tensl
    \def\bf{\fam\bffam \fourteenbf\f@ntkey=6 }\textfont\bffam=\fourteenbf
    \scriptfont\bffam=\tenbf	 \scriptscriptfont\bffam=\sevenbf
    \def\tt{\fam\ttfam \twelvett \f@ntkey=7 }\textfont\ttfam=\twelvett
    \h@big=11.9\p@ \h@Big=16.1\p@ \h@bigg=20.3\p@ \h@Bigg=24.5\p@
    \def\caps{\fam\cpfam \twelvecp \f@ntkey=8 }\textfont\cpfam=\twelvecp
    \setbox\strutbox=\hbox{\vrule height 12pt depth 5pt width\z@}\relax
    \samef@nt}
%%%%%%%%%%%%%%%%%%%%%%%%%%%%%%%%%%%%%%%%%%%%%%%%%%%%%%%%%%%%%%%%%%%%%%%%%
% scriptscriptfont size changed to 7pt
% Feb.10,1988       K.Hikasa
%%%%%%%%%%%%%%%%%%%%%%%%%%%%%%%%%%%%%%%%%%%%%%%%%%%%%%%%%%%%%%%%%%%%%%%%%
\def\twelvepoint{\relax
    \textfont0=\twelverm	  \scriptfont0=\ninerm
    \scriptscriptfont0=\sevenrm
     \def\rm{\fam0 \twelverm \f@ntkey=0 }\relax
    \textfont1=\twelvei		  \scriptfont1=\ninei
    \scriptscriptfont1=\seveni
     \def\oldstyle{\fam1 \twelvei\f@ntkey=1 }\relax
    \textfont2=\twelvesy	  \scriptfont2=\ninesy
    \scriptscriptfont2=\sevensy
    \textfont3=\twelveex	  \scriptfont3=\twelveex
    \scriptscriptfont3=\twelveex
    \def\it{\fam\itfam \twelveit \f@ntkey=4 }\textfont\itfam=\twelveit
    \def\sl{\fam\slfam \twelvesl \f@ntkey=5 }\textfont\slfam=\twelvesl
    \scriptfont\slfam=\ninesl
    \def\bf{\fam\bffam \twelvebf \f@ntkey=6 }\textfont\bffam=\twelvebf
    \scriptfont\bffam=\ninebf	  \scriptscriptfont\bffam=\sevenbf
    \def\tt{\fam\ttfam \twelvett \f@ntkey=7 }\textfont\ttfam=\twelvett
    \h@big=10.2\p@ \h@Big=13.8\p@ \h@bigg=17.4\p@ \h@Bigg=21.0\p@
    \def\caps{\fam\cpfam \twelvecp \f@ntkey=8 }\textfont\cpfam=\twelvecp
    \setbox\strutbox=\hbox{\vrule height 10pt depth 4pt width\z@}\relax
    \samef@nt}
\def\tenpoint{\relax
    \textfont0=\tenrm	       \scriptfont0=\sevenrm
    \scriptscriptfont0=\fiverm
    \def\rm{\fam0 \tenrm \f@ntkey=0 }\relax
    \textfont1=\teni	       \scriptfont1=\seveni
    \scriptscriptfont1=\fivei
    \def\oldstyle{\fam1 \teni \f@ntkey=1 }\relax
    \textfont2=\tensy	       \scriptfont2=\sevensy
    \scriptscriptfont2=\fivesy
    \textfont3=\tenex	       \scriptfont3=\tenex
    \scriptscriptfont3=\tenex
    \def\it{\fam\itfam \tenit \f@ntkey=4 }\textfont\itfam=\tenit
    \def\sl{\fam\slfam \tensl \f@ntkey=5 }\textfont\slfam=\tensl
    \def\bf{\fam\bffam \tenbf \f@ntkey=6 }\textfont\bffam=\tenbf
    \scriptfont\bffam=\sevenbf	   \scriptscriptfont\bffam=\fivebf
    \def\tt{\fam\ttfam \tentt \f@ntkey=7 }\textfont\ttfam=\tentt
    \def\caps{\fam\cpfam \tencp \f@ntkey=8 }\textfont\cpfam=\tencp
    \h@big=8.5\p@ \h@Big=11.5\p@ \h@bigg=14.5\p@ \h@Bigg=17.5\p@
    \setbox\strutbox=\hbox{\vrule height 8.5pt depth 3.5pt width\z@}\relax
    \samef@nt}
%%%%%%%%%%%%%%%%%%%%%%%%%%%%%%%%%%%%%%%%%%%%%%%%%%%%%%%%%%%%%%%%%%%%%%%%
%   Next redefine \big \Big \bigg and \Bigg to work with all fonts
%%%%%%%%%%%%%%%%%%%%%%%%%%%%%%%%%%%%%%%%%%%%%%%%%%%%%%%%%%%%%%%%%%%%%%%%
\newdimen\h@big  \h@big=8.5\p@
\newdimen\h@Big  \h@Big=11.5\p@
\newdimen\h@bigg  \h@bigg=14.5\p@
\newdimen\h@Bigg  \h@Bigg=17.5\p@
\def\big#1{{\hbox{$\left#1\vbox to\h@big{}\right.\n@space$}}}
\def\Big#1{{\hbox{$\left#1\vbox to\h@Big{}\right.\n@space$}}}
\def\bigg#1{{\hbox{$\left#1\vbox to\h@bigg{}\right.\n@space$}}}
\def\Bigg#1{{\hbox{$\left#1\vbox to\h@Bigg{}\right.\n@space$}}}
%%%%%%%%%%%%%%%%%%%%%%%%%%%%%%%%%%%%%%%%%%%%%%%%%%%%%%%%%%%%%%%%%%%%%%%%
%   Next, I define basic spacing parameters.
%
\normalbaselineskip = 20pt plus 0.2pt minus 0.1pt
\normallineskip = 1.5pt plus 0.1pt minus 0.1pt
\normallineskiplimit = 1.5pt
\newskip\normaldisplayskip
\normaldisplayskip = 20pt plus 5pt minus 10pt
\newskip\normaldispshortskip
\normaldispshortskip = 6pt plus 5pt
\newskip\normalparskip
\normalparskip = 6pt plus 2pt minus 1pt
\newskip\skipregister
\skipregister = 5pt plus 2pt minus 1.5pt
\newif\ifsingl@	   \newif\ifdoubl@
\newif\iftwelv@	   \twelv@true
\def\singlespace{\singl@true\doubl@false\spaces@t}
\def\doublespace{\singl@false\doubl@true\spaces@t}
\def\normalspace{\singl@false\doubl@false\spaces@t}
\def\Tenpoint{\tenpoint\twelv@false\spaces@t}
\def\Twelvepoint{\twelvepoint\twelv@true\spaces@t}
\def\spaces@t{\relax
 \iftwelv@\ifsingl@\subspaces@t3:4;\else\subspaces@t1:1;\fi
 \else\ifsingl@\subspaces@t3:5;\else\subspaces@t4:5;\fi\fi
 \ifdoubl@\multiply\baselineskip by 5 \divide\baselineskip by 4 \fi}
\def\subspaces@t#1:#2;{\baselineskip=\normalbaselineskip
\multiply\baselineskip by #1\divide\baselineskip by #2%
\lineskip = \normallineskip
\multiply\lineskip by #1\divide\lineskip by #2%
\lineskiplimit = \normallineskiplimit
\multiply\lineskiplimit by #1\divide\lineskiplimit by #2%
\parskip = \normalparskip
\multiply\parskip by #1\divide\parskip by #2%
\abovedisplayskip = \normaldisplayskip
\multiply\abovedisplayskip by #1\divide\abovedisplayskip by #2%
\belowdisplayskip = \abovedisplayskip
\abovedisplayshortskip = \normaldispshortskip
\multiply\abovedisplayshortskip by #1%
\divide\abovedisplayshortskip by #2%
\belowdisplayshortskip = \abovedisplayshortskip
\advance\belowdisplayshortskip by \belowdisplayskip
\divide\belowdisplayshortskip by 2
\smallskipamount = \skipregister
\multiply\smallskipamount by #1\divide\smallskipamount by #2%
\medskipamount = \smallskipamount \multiply\medskipamount by 2
\bigskipamount = \smallskipamount \multiply\bigskipamount by 4 }
\def\normalbaselines{ \baselineskip=\normalbaselineskip%
   \lineskip=\normallineskip \lineskiplimit=\normallineskip%
   \iftwelv@\else \multiply\baselineskip by 4 \divide\baselineskip by 5%
     \multiply\lineskiplimit by 4 \divide\lineskiplimit by 5%
     \multiply\lineskip by 4 \divide\lineskip by 5 \fi }
\Twelvepoint  % That's the default
\interlinepenalty=50
\interfootnotelinepenalty=5000
\predisplaypenalty=9000
\postdisplaypenalty=500
\hfuzz=1pt
\vfuzz=0.2pt
%%%%%%%%%%%%%%%%%%%%%%%%%%%%%%%%%%%%%%%%%%%%%%%%%%%%%%%%%%%%%%%%%%%%%%%%
%   Next, I define output routines, footnotes & related stuff.
%
\def\pagecontents{%
   \ifvoid\topins\else\unvbox\topins\vskip\skip\topins\fi
   \dimen@ = \dp255 \unvbox255
   \ifvoid\footins\else\vskip\skip\footins\footrule\unvbox\footins\fi
   \ifr@ggedbottom \kern-\dimen@ \vfil \fi }
\def\makeheadline{\vbox to 0pt{ \skip@=\topskip
      \advance\skip@ by -12pt \advance\skip@ by -2\normalbaselineskip
      \vskip\skip@ \line{\vbox to 12pt{}\the\headline} \vss
      }\nointerlineskip}
\def\makefootline{\baselineskip = 1.5\normalbaselineskip
		 \line{\the\footline}}
\newif\iffrontpage
\newif\ifletterstyle
\newif\ifp@genum
\def\nopagenumbers{\p@genumfalse}
\def\pagenumbers{\p@genumtrue}
\pagenumbers
\newtoks\paperheadline
\newtoks\letterheadline
\newtoks\letterfrontheadline
\newtoks\lettermainheadline
\newtoks\paperfootline
\newtoks\letterfootline
\newtoks\date
\newtoks\Month
\footline={\ifletterstyle\the\letterfootline\else\the\paperfootline\fi}
\paperfootline={\hss\iffrontpage\else\ifp@genum\tenrm\folio\hss\fi\fi}
\letterfootline={\hfil}
\headline={\ifletterstyle\the\letterheadline\else\the\paperheadline\fi}
\paperheadline={\hfil}
\letterheadline{\iffrontpage\the\letterfrontheadline
     \else\the\lettermainheadline\fi}
\lettermainheadline={\rm\ifp@genum page \ \folio\fi\hfil\the\date}
\def\monthname{\relax\ifcase\month 0/\or January\or February\or
   March\or April\or May\or June\or July\or August\or September\or
   October\or November\or December\else\number\month/\fi}
\date={\monthname\ \number\day, \number\year}
\Month={\monthname\ \number\year}
\countdef\pagenumber=1  \pagenumber=1
\def\advancepageno{\global\advance\pageno by 1
   \ifnum\pagenumber<0 \global\advance\pagenumber by -1
    \else\global\advance\pagenumber by 1 \fi \global\frontpagefalse }
\def\folio{\ifnum\pagenumber<0 \romannumeral-\pagenumber
	   \else \number\pagenumber \fi }
\def\footrule{\dimen@=\prevdepth\nointerlineskip
   \vbox to 0pt{\vskip -0.25\baselineskip \hrule width 0.35\hsize \vss}
   \prevdepth=\dimen@ }
\newtoks\foottokens
\foottokens={\Tenpoint\singlespace}
\newdimen\footindent
\footindent=24pt
\def\vfootnote#1{\insert\footins\bgroup  \the\foottokens
   \interlinepenalty=\interfootnotelinepenalty \floatingpenalty=20000
   \splittopskip=\ht\strutbox \boxmaxdepth=\dp\strutbox
   \leftskip=\footindent \rightskip=\z@skip
   \parindent=0.5\footindent \parfillskip=0pt plus 1fil
   \spaceskip=\z@skip \xspaceskip=\z@skip
   \Textindent{$ #1 $}\footstrut\futurelet\next\fo@t}
\def\Textindent#1{\noindent\llap{#1\enspace}\ignorespaces}
\def\footnote#1{\attach{#1}\vfootnote{#1}}

\let\footsymbol=\star
\newcount\lastf@@t	     \lastf@@t=-1
\newcount\footsymbolcount    \footsymbolcount=0
\newif\ifPhysRev
\def\footsymbolgen{\relax \ifPhysRev \iffrontpage \NPsymbolgen\else
      \PRsymbolgen\fi \else \NPsymbolgen\fi
   \global\lastf@@t=\pageno \footsymbol }
\def\NPsymbolgen{\ifnum\footsymbolcount<0 \global\footsymbolcount=0\fi
   {\iffrontpage \else \advance\lastf@@t by 1 \fi
    \ifnum\lastf@@t<\pageno \global\footsymbolcount=0
     \else \global\advance\footsymbolcount by 1 \fi }
   \ifcase\footsymbolcount \fd@f\star\or \fd@f\dagger\or \fd@f\ast\or
    \fd@f\ddagger\or \fd@f\natural\or \fd@f\diamond\or \fd@f\bullet\or
    \fd@f\nabla\else \fd@f\dagger\global\footsymbolcount=0 \fi }
\def\fd@f#1{\xdef\footsymbol{#1}}
\def\PRsymbolgen{\ifnum\footsymbolcount>0 \global\footsymbolcount=0\fi
      \global\advance\footsymbolcount by -1
      \xdef\footsymbol{\sharp\number-\footsymbolcount} }
\def\space@ver#1{\let\@sf=\empty \ifmmode #1\else \ifhmode
   \edef\@sf{\spacefactor=\the\spacefactor}\unskip${}#1$\relax\fi\fi}
\def\attach#1{\space@ver{\strut^{\mkern 2mu #1} }\@sf\ }
%%%%%%%%%%%%%%%%%%%%%%%%%%%%%%%%%%%%%%%%%%%%%%%%%%%%%%%%%%%%%%%%%%%%%%%%
%   Here come chapter, section, subsection & appendix macros.
%
\newcount\chapternumber	     \chapternumber=0
\newcount\sectionnumber	     \sectionnumber=0
\newcount\equanumber	     \equanumber=0
\let\chapterlabel=\relax
\newtoks\chapterstyle	     \chapterstyle={\Number}
\newskip\chapterskip	     \chapterskip=\bigskipamount
\newskip\sectionskip	     \sectionskip=\medskipamount
\newskip\headskip	     \headskip=8pt plus 3pt minus 3pt
\newdimen\chapterminspace    \chapterminspace=15pc
\newdimen\sectionminspace    \sectionminspace=10pc
\newdimen\referenceminspace  \referenceminspace=25pc
\def\chapterreset{\global\advance\chapternumber by 1
   \ifnum\equanumber<0 \else\global\equanumber=0\fi
   \sectionnumber=0 \makel@bel}
\def\makel@bel{\xdef\chapterlabel{%
\the\chapterstyle{\the\chapternumber}.}}
\def\sectionlabel{\number\sectionnumber \quad }
\def\alphabetic#1{\count255='140 \advance\count255 by #1\char\count255}
\def\Alphabetic#1{\count255='100 \advance\count255 by #1\char\count255}
\def\Roman#1{\uppercase\expandafter{\romannumeral #1}}
\def\roman#1{\romannumeral #1}
\def\Number#1{\number #1}
\def\unnumberedchapters{\let\makel@bel=\relax \let\chapterlabel=\relax
\let\sectionlabel=\relax \equanumber=-1 }
\def\titlestyle#1{\par\begingroup \interlinepenalty=9999
     \leftskip=0.02\hsize plus 0.23\hsize minus 0.02\hsize
     \rightskip=\leftskip \parfillskip=0pt
     \hyphenpenalty=9000 \exhyphenpenalty=9000
     \tolerance=9999 \pretolerance=9000
     \spaceskip=0.333em \xspaceskip=0.5em
     \iftwelv@\fourteenpoint\else\twelvepoint\fi
   \noindent #1\par\endgroup }
\def\spacecheck#1{\dimen@=\pagegoal\advance\dimen@ by -\pagetotal
   \ifdim\dimen@<#1 \ifdim\dimen@>0pt \vfil\break \fi\fi}
\def\chapter#1{\par \penalty-300 \vskip\chapterskip
   \spacecheck\chapterminspace
   \chapterreset \titlestyle{\chapterlabel \ #1}
   \nobreak\vskip\headskip \penalty 30000
   \wlog{\string\chapter\ \chapterlabel} }

\def\section#1{\par \ifnum\the\lastpenalty=30000\else
   \penalty-200\vskip\sectionskip \spacecheck\sectionminspace\fi
   \wlog{\string\section\ \chapterlabel \the\sectionnumber}
   \global\advance\sectionnumber by 1  \noindent
   {\caps\enspace\chapterlabel \sectionlabel #1}\par
   \nobreak\vskip\headskip \penalty 30000 }
\def\subsection#1{\par
   \ifnum\the\lastpenalty=30000\else \penalty-100\smallskip \fi
   \noindent\undertext{#1}\enspace \vadjust{\penalty5000}}

\def\undertext#1{\vtop{\hbox{#1}\kern 1pt \hrule}}
\def\ack{\par\penalty-100\medskip \spacecheck\sectionminspace
   \line{\fourteenrm\hfil ACKNOWLEDGMENTS\hfil}\nobreak\vskip\headskip }
\def\APPENDIX#1#2{\par\penalty-300\vskip\chapterskip
   \spacecheck\chapterminspace \chapterreset \xdef\chapterlabel{#1}
   \titlestyle{APPENDIX #2} \nobreak\vskip\headskip \penalty 30000
   \wlog{\string\Appendix\ \chapterlabel} }
\def\Appendix#1{\APPENDIX{#1}{#1}}
\def\appendix{\APPENDIX{A}{}}
%%%%%%%%%%%%%%%%%%%%%%%%%%%%%%%%%%%%%%%%%%%%%%%%%%%%%%%%%%%%%%%%%%%%%%%%
%   Here come macros for equation numbering.
%
\def\eqname#1{\relax \ifnum\equanumber<0
     \xdef#1{{\rm(\number-\equanumber)}}\global\advance\equanumber by -1
    \else \global\advance\equanumber by 1
      \xdef#1{{\rm(\chapterlabel \number\equanumber)}} \fi}
\def\eq{\eqname\?\?}
\def\eqn#1{\eqno\eqname{#1}#1}

\def\eqinsert#1{\noalign{\dimen@=\prevdepth \nointerlineskip
   \setbox0=\hbox to\displaywidth{\hfil #1}
   \vbox to 0pt{\vss\hbox{$\!\box0\!$}\kern-0.5\baselineskip}
   \prevdepth=\dimen@}}

%%%%%%%%%%%%%%%%%%%%%%%%%%%%%%%%%%%%%%%%%%%%%%%%%%%%%%%%%%%%%%%%%%%%%%%%
%   Here come items and lists
%
\def\GENITEM#1;#2{\par \hangafter=0 \hangindent=#1
    \Textindent{$ #2 $}\ignorespaces}
\outer\def\newitem#1=#2;{\gdef#1{\GENITEM #2;}}
\newdimen\itemsize		  \itemsize=30pt
\newitem\item=1\itemsize;
\newitem\sitem=1.75\itemsize;	  
\newitem\ssitem=2.5\itemsize;	  
\outer\def\newlist#1=#2&#3&#4;{\toks0={#2}\toks1={#3}%
   \count255=\escapechar \escapechar=-1
   \alloc@0\list\countdef\insc@unt\listcount	 \listcount=0
   \edef#1{\par
      \countdef\listcount=\the\allocationnumber
      \advance\listcount by 1
      \hangafter=0 \hangindent=#4
      \Textindent{\the\toks0{\listcount}\the\toks1}}
   \expandafter\expandafter\expandafter
    \edef\c@t#1{begin}{\par
      \countdef\listcount=\the\allocationnumber \listcount=1
      \hangafter=0 \hangindent=#4
      \Textindent{\the\toks0{\listcount}\the\toks1}}
   \expandafter\expandafter\expandafter
    \edef\c@t#1{con}{\par \hangafter=0 \hangindent=#4 \noindent}
   \escapechar=\count255}
\def\c@t#1#2{\csname\string#1#2\endcsname}
\newlist\point=\Number&.&1.0\itemsize;
\newlist\subpoint=(\alphabetic&)&1.75\itemsize;
\newlist\subsubpoint=(\roman&)&2.5\itemsize;

%%%%%%%%%%%%%%%%%%%%%%%%%%%%%%%%%%%%%%%%%%%%%%%%%%%%%%%%%%%%%%%%%%%%%%%%
%   Here come macros for references, figures & tables.
%
\newcount\referencecount     \referencecount=0
\newif\ifreferenceopen	     \newwrite\referencewrite
\newtoks\rw@toks
\def\NPrefmark#1{\attach{\scriptscriptstyle [ #1 ] }}
\let\PRrefmark=\attach
\def\refmark#1{\relax\ifPhysRev\PRrefmark{#1}\else\NPrefmark{#1}\fi}
\def\refend{\refmark{\number\referencecount}}
\newcount\lastrefsbegincount \lastrefsbegincount=0
\def\refsend{\refmark{\count255=\referencecount
   \advance\count255 by-\lastrefsbegincount
   \ifcase\count255 \number\referencecount
   \or \number\lastrefsbegincount,\number\referencecount
   \else \number\lastrefsbegincount-\number\referencecount \fi}}
\def\refch@ck{\chardef\rw@write=\referencewrite
   \ifreferenceopen \else \referenceopentrue
   \immediate\openout\referencewrite=reference.aux \fi}
% In \obeyendofline, we say `\let^^M=\relax
{\catcode`\^^M=\active % these lines must end with %
  \gdef\obeyendofline{\catcode`\^^M\active \let^^M\ }}%
% In \ignoreendofline, we say `\let^^M=\relax
{\catcode`\^^M=\active % these lines must end with %
  \gdef\ignoreendofline{\catcode`\^^M=5}}
{\obeyendofline\gdef\rw@start#1{\def\t@st{#1} \ifx\t@st\blankend%
\endgroup \@sf \relax \else \ifx\t@st\bl@nkend \endgroup \@sf \relax%
\else \rw@begin#1
\backtotext
\fi \fi } }
{\obeyendofline\gdef\rw@begin#1
{\def\n@xt{#1}\rw@toks={#1}\relax%
\rw@next}}
\def\blankend{}
{\obeylines\gdef\bl@nkend{
}}
\newif\iffirstrefline  \firstreflinetrue
\def\rwr@teswitch{\ifx\n@xt\blankend \let\n@xt=\rw@begin %
 \else\iffirstrefline \global\firstreflinefalse%
\immediate\write\rw@write{\noexpand\obeyendofline \the\rw@toks}%
\let\n@xt=\rw@begin%
      \else\ifx\n@xt\rw@@d \def\n@xt{\immediate\write\rw@write{%
	\noexpand\ignoreendofline}\endgroup \@sf}%
	     \else \immediate\write\rw@write{\the\rw@toks}%
	     \let\n@xt=\rw@begin\fi\fi \fi}
\def\rw@next{\rwr@teswitch\n@xt}
\def\rw@@d{\backtotext} \let\rw@end=\relax
\let\backtotext=\relax

\newdimen\refindent	\refindent=30pt
\def\refitem#1{\par \hangafter=0 \hangindent=\refindent \Textindent{#1}}
\def\REFNUM#1{\space@ver{}\refch@ck \firstreflinetrue%
 \global\advance\referencecount by 1 \xdef#1{\the\referencecount}}
\def\refnum#1{\space@ver{}\refch@ck \firstreflinetrue%
 \global\advance\referencecount by 1 \xdef#1{\the\referencecount}\refend}
\def\REF#1{\REFNUM#1%
 \immediate\write\referencewrite{%
 \noexpand\refitem{#1.}}%
\begingroup\obeyendofline\rw@start}
\def\ref{\refnum\?%
 \immediate\write\referencewrite{\noexpand\refitem{\?.}}%
\begingroup\obeyendofline\rw@start}
\def\Ref#1{\refnum#1%
 \immediate\write\referencewrite{\noexpand\refitem{#1.}}%
\begingroup\obeyendofline\rw@start}
\def\REFS#1{\REFNUM#1\global\lastrefsbegincount=\referencecount
\immediate\write\referencewrite{\noexpand\refitem{#1.}}%
\begingroup\obeyendofline\rw@start}

\newcount\figurecount	  \figurecount=0
\newif\iffigureopen	  \newwrite\figurewrite
\def\figch@ck{\chardef\rw@write=\figurewrite \iffigureopen\else
   \immediate\openout\figurewrite=figures.aux
   \figureopentrue\fi}
\def\FIGNUM#1{\space@ver{}\figch@ck \firstreflinetrue%
 \global\advance\figurecount by 1 \xdef#1{\the\figurecount}}
\def\FIG#1{\FIGNUM#1
   \immediate\write\figurewrite{\noexpand\refitem{#1.}}%
   \begingroup\obeyendofline\rw@start}
\def\FIGFIG#1{\FIGNUM#1
   \immediate\write\figurewrite{\noexpand\refitem{Fig.#1.}}%
   \begingroup\obeyendofline\rw@start}
\def\figout{\par \penalty-400 \vskip\chapterskip
   \spacecheck\referenceminspace \immediate\closeout\figurewrite
   \figureopenfalse
   \line{\fourteenrm\hfil FIGURE CAPTIONS\hfil}\vskip\headskip
   \input figures.aux
   }

\def\fig{\FIGNUM\? fig.~\?%
\immediate\write\figurewrite{\noexpand\refitem{\?.}}%
\begingroup\obeyendofline\rw@start}
\def\figure{\FIGNUM\? figure~\?
   \immediate\write\figurewrite{\noexpand\refitem{\?.}}%
   \begingroup\obeyendofline\rw@start}
\def\Fig{\FIGNUM\? Fig.~\?%
\immediate\write\figurewrite{\noexpand\refitem{\?.}}%
\begingroup\obeyendofline\rw@start}
\def\Figure{\FIGNUM\? Figure~\?%
\immediate\write\figurewrite{\noexpand\refitem{\?.}}%
\begingroup\obeyendofline\rw@start}
\newcount\tablecount	 \tablecount=0
\newif\iftableopen	 \newwrite\tablewrite
\def\tabch@ck{\chardef\rw@write=\tablewrite \iftableopen\else
   \immediate\openout\tablewrite=tables.aux
   \tableopentrue\fi}
\def\TABNUM#1{\space@ver{}\tabch@ck \firstreflinetrue%
 \global\advance\tablecount by 1 \xdef#1{\the\tablecount}}
\def\TABLE#1{\TABNUM#1
   \immediate\write\tablewrite{\noexpand\refitem{#1.}}%
   \begingroup\obeyendofline\rw@start}
\def\Table{\TABNUM\? Table~\?%
\immediate\write\tablewrite{\noexpand\refitem{\?.}}%
\begingroup\obeyendofline\rw@start}
\def\tabout{\par \penalty-400 \vskip\chapterskip
   \spacecheck\referenceminspace \immediate\closeout\tablewrite
   \tableopenfalse
   \line{\fourteenrm\hfil TABLE CAPTIONS\hfil}\vskip\headskip
   \input tables.aux
   }
%%%%%%%%%%%%%%%%%%%%%%%%%%%%%%%%%%%%%%%%%%%%%%%%%%%%%%%%%%%%%%%%%%%%%%%%
%   Here come macros for memos & letters.
%
\def\masterreset{\global\pagenumber=1 \global\chapternumber=0
   \global\equanumber=0 \global\sectionnumber=0
   \global\referencecount=0 \global\figurecount=0 \global\tablecount=0 }
\def\FRONTPAGE{\ifvoid255\else\vfill\penalty-2000\fi
      \masterreset\global\frontpagetrue
      \global\lastf@@t=0 \global\footsymbolcount=0}

\def\paperstyle{\letterstylefalse\normalspace\papersize}
\def\letterstyle{\letterstyletrue\singlespace\lettersize}
\def\papersize{\hsize=35pc\vsize=50pc\hoffset=1pc\voffset=6pc
		\skip\footins=\bigskipamount}
\def\lettersize{\hsize=6in\vsize=8.5in\hoffset=0.33in\voffset=1in
   \skip\footins=\smallskipamount \multiply\skip\footins by 3 }
\paperstyle   %  This is the default
%
% % % % % % % % % % % % % % % % % % % % % % % % % % % % % % % % % % % %
%
\def\MEMO{\letterstyle\FRONTPAGE \letterfrontheadline={\hfil}
    \line{\quad\fourteenrm KEK MEMORANDUM\hfil\twelverm\the\date\quad}
    \medskip \memod@f}

\def\memit@m#1{\smallskip \hangafter=0 \hangindent=1in
      \Textindent{\caps #1}}
\def\memod@f{\xdef\to{\memit@m{To:}}\xdef\from{\memit@m{From:}}%
     \xdef\topic{\memit@m{Topic:}}\xdef\subject{\memit@m{Subject:}}%
     \xdef\rule{\bigskip\hrule height 1pt\bigskip}}
\memod@f

\def\nohead{
 \def\letters{\letterstyle \letterfrontheadline={\hfil}}
 \def\letter{\FRONTPAGE\BLANKHEAD\addressee}%for use of a headed letter paper
}
\nohead

\def\BLANKHEAD{\hrule height 0pt depth 0pt \vskip 1cm plus .5cm minus .5cm}
\newskip\lettertopfil
\lettertopfil = 0pt plus 1.5in minus 0pt
\newskip\letterbottomfil
\letterbottomfil = 0pt plus 2.3in minus 0pt
\newskip\spskip \setbox0\hbox{\ } \spskip=-1\wd0
\def\addressee#1{\medskip \line{\hskip 0.5\hsize \hbox{\the\date}\hfil}
   \bigskip
   \vskip\lettertopfil
   \ialign to\hsize{\strut ##\hfil\tabskip 0pt plus \hsize \cr #1\crcr}
   \medskip\noindent\hskip\spskip}
\newskip\signatureskip	     \signatureskip=40pt
\def\signed#1{\par \penalty 9000 \bigskip \dt@pfalse
  \everycr={\noalign{\ifdt@p\vskip\signatureskip\global\dt@pfalse\fi}}
  \setbox0=\vbox{\singlespace \halign{\tabskip 0pt \strut ##\hfil\cr
   \noalign{\global\dt@ptrue}#1\crcr}}
  \line{\hskip 0.5\hsize minus 0.5\hsize \box0\hfil} \medskip }

\def\endletter{\ifnum\pagenumber=1 \vskip\letterbottomfil\supereject
\else \vfil\supereject \fi}
\newbox\letterb@x
\def\lettertext{\par\unvcopy\letterb@x\par}
\def\multiletter{\setbox\letterb@x=\vbox\bgroup
      \everypar{\vrule height 1\baselineskip depth 0pt width 0pt }
      \singlespace \topskip=\baselineskip }
\def\letterend{\par\egroup}
%%%%%%%%%%%%%%%%%%%%%%%%%%%%%%%%%%%%%%%%%%%%%%%%%%%%%%%%%%%%%%%%%%%%%%%
%   Here come macros for title pages.
%%%%%%%%%%%%%%%%%%%%%%%%%%%%%%%%%%%%%%%%%%%%%%%%%%%%%%%%%%%%%%%%%%%%%%
% Macros for pubblock rewritten by H. Mawatari, S. Ohta, and K. Hikasa
%%%%%%%%%%%%%%%%%%%%%%%%%%%%%%%%%%%%%%%%%%%%%%%%%%%%%%%%%%%%%%%%%%%%%%
\newskip\frontpageskip
\newtoks\pubtype
\newtoks\Pubnum \newtoks\pubnum
\newtoks\Thnum \newtoks\thnum
\newtoks\s@condpubnum \newtoks\th@rdpubnum
\newif\ifs@cond \s@condfalse
\newif\ifth@rd \th@rdfalse
\newif\ifp@bblock  \p@bblocktrue
\newcount\Year
\def\Yearset{\Year=\year \advance\Year by -1900
 \ifnum\month<4 \advance\Year by -1 \fi}
\def\PH@SR@V{\doubl@true \baselineskip=24.1pt plus 0.2pt minus 0.1pt
	     \parskip= 3pt plus 2pt minus 1pt }
\def\PHYSREV{\paperstyle\PhysRevtrue\PH@SR@V}
\def\titlepage{\Yearset\FRONTPAGE\paperstyle\ifPhysRev\PH@SR@V\fi
   \ifp@bblock\p@bblock\fi}
\def\nopubblock{\p@bblockfalse}
\def\endpage{\vfil\break}
\frontpageskip=1\medskipamount plus .5fil
\pubtype={\tensl Preliminary Version}
\newtoks\publevel
\publevel={Preprint}   % The alternatives are Internal and Preprint
\Pubnum={KEK \the\publevel\ \the\Year--\the\pubnum }
\pubnum={ }
\Thnum={KEK--TH--\the\thnum }
\thnum={ }
\def\secondpubnum#1{\s@condtrue\s@condpubnum={#1}}
\def\thirdpubnum#1{\th@rdtrue\th@rdpubnum={#1}}
\def\p@bblock{\begingroup \tabskip=\hsize minus \hsize
   \baselineskip=1.5\ht\strutbox \topspace-2\baselineskip
   \halign to\hsize{\strut ##\hfil\tabskip=0pt\crcr
   \the\Thnum\cr \the\Pubnum\cr
   \ifs@cond \the\s@condpubnum\cr\fi
   \ifth@rd \the\th@rdpubnum\cr\fi
   \the\Month \cr}\endgroup}
\def\title#1{\hrule height0pt depth0pt
   \vskip\frontpageskip \titlestyle{#1} \vskip\headskip }
%%%%%%%%%%%%%%%%%%%%%%%%%%%%%%%%%%%%%%%%%%%%%%%%%%%%%%%%%%%%%%%%%%%%%%%%%%%
\def\author#1{\vskip\frontpageskip\titlestyle{\twelvecp #1}\nobreak}

\def\address#1{\par\kern 5pt\titlestyle{\twelvepoint\it #1}}
\def\andaddress{\par\kern 5pt \centerline{\sl and} \address}

\def\abstract{\vskip\frontpageskip\centerline{\fourteenrm ABSTRACT}
	      \vskip\headskip }

%%%%%%%%%%%%%%%%%%%%%%%%%%%%%%%%%%%%%%%%%%%%%%%%%%%%%%%%%%%%%%%%%%%%%%%%
%   Miscellaneous macros
%

\def\\{\relax\ifmmode\backslash\else$\backslash$\fi}
\def\globaleqnumbers{\relax\if\equanumber<0\else\global\equanumber=-1\fi}

\def\journal#1&#2(#3){\unskip, \sl #1~\bf #2 \rm (19#3) }

\def\topspace{\hrule height 0pt depth 0pt \vskip}

\let\int=\intop		
\def\prop{\mathrel{{\mathchoice{\pr@p\scriptstyle}{\pr@p\scriptstyle}{
		\pr@p\scriptscriptstyle}{\pr@p\scriptscriptstyle} }}}
\def\pr@p#1{\setbox0=\hbox{$\cal #1 \char'103$}
   \hbox{$\cal #1 \char'117$\kern-.4\wd0\box0}}
\def\lsim{\mathrel{\mathpalette\@versim<}}
\def\gsim{\mathrel{\mathpalette\@versim>}}
\def\@versim#1#2{\lower0.2ex\vbox{\baselineskip\z@skip\lineskip\z@skip
  \lineskiplimit\z@\ialign{$\m@th#1\hfil##\hfil$\crcr#2\crcr\sim\crcr}}}
% % % % % % % % % % % % % % % % % % % % % % % % % % % % % % % % % % % %
%   Finally, some bug fixings.
%
\let\sec@nt=\sec
\def\sec{\relax\ifmmode\let\n@xt=\sec@nt\else\let\n@xt\section\fi\n@xt}
\def\obsolete#1{\message{Macro \string #1 is obsolete.}}
\def\firstsec#1{\obsolete\firstsec \section{#1}}
\def\firstsubsec#1{\obsolete\firstsubsec \subsection{#1}}
\def\thispage#1{\obsolete\thispage \global\pagenumber=#1\frontpagefalse}
\def\thischapter#1{\obsolete\thischapter \global\chapternumber=#1}
\def\nextequation#1{\obsolete\nextequation \global\equanumber=#1
   \ifnum\the\equanumber>0 \global\advance\equanumber by 1 \fi}
\def\BOXITEM{\afterassigment\B@XITEM\setbox0=}
\def\B@XITEM{\par\hangindent\wd0 \noindent\box0 }

%%%%%%%%%%%%%%%%%%%%%%%%%%%%%%%%%%%%%%%%%%%%%%%%%%%%%%%%%%%%%%%%%%%%%%%%
%   That's about it
%
\catcode`@=12 % at signs are no longer letters
\message{ by V.K.}
\everyjob{\input myphyx }
%%%%%%%%%%%%%%%%%%%%%%%%%%%%%%%%%%%%%%%%%%%%%%%%%%%%%%%%%%%%%%%%%%%%%%%%
%  Addition by K. Hikasa, April 1, 1987                                 %
%%%%%%%%%%%%%%%%%%%%%%%%%%%%%%%%%%%%%%%%%%%%%%%%%%%%%%%%%%%%%%%%%%%%%%%%%
   %cf. The TeX Book p.~356
%%%%%%%%%%%%%%%%%%%%%%%%%%%%%%%%%%%%%%%%%%%%%%%%%%%%%%%%%%%%%%%%%%%%%%%%%
%  This is a macro package ELEVENPT.TEX for the use of 11 point letters
%             By K. Hikasa, March 16, 1987
%             bug fixed     July 29, 1989
%%%%%%%%%%%%%%%%%%%%%%%%%%%%%%%%%%%%%%%%%%%%%%%%%%%%%%%%%%%%%%%%%%%%%%%%%
\font\elevenrm=cmr10 scaled\magstephalf
\font\elevenbf=cmbx10 scaled\magstephalf
\font\eleveni=cmmi10 scaled\magstephalf \skewchar\eleveni='177
\font\elevensy=cmsy10 scaled\magstephalf \skewchar\elevensy='60
\font\elevenex=cmex10 scaled\magstephalf
\font\elevensl=cmsl10 scaled\magstephalf
\font\elevenit=cmti10 scaled\magstephalf
\font\eleventt=cmtt10 scaled\magstephalf
\font\elevencp=cmcsc10 scaled\magstephalf
\catcode\lq@=11 %
\def\elevenpoint{\relax
 \textfont0=\elevenrm \scriptfont0=\ninerm \scriptscriptfont0=\sixrm
 \def\rm{\fam0 \elevenrm \f@ntkey=0}\relax
 \textfont1=\eleveni \scriptfont1=\ninei \scriptscriptfont1=\sixi
 \def\oldstyle{\fam1 \eleveni\f@ntkey=1}\relax
 \textfont2=\elevensy \scriptfont2=\ninesy \scriptscriptfont2=\sixsy
 \textfont3=\elevenex \scriptfont3=\elevenex \scriptscriptfont3=\elevenex
 \def\it{\fam\itfam \elevenit \f@ntkey=4 }\textfont\itfam=\elevenit
 \def\sl{\fam\slfam \elevensl \f@ntkey=5 }\textfont\slfam=\elevensl
 \scriptfont\slfam=\ninesl
 \def\bf{\fam\bffam \elevenbf \f@ntkey=6 }\textfont\bffam=\elevenbf
 \scriptfont\bffam=\ninebf \scriptscriptfont\bffam=\sixbf
 \def\tt{\fam\ttfam \eleventt \f@ntkey=7 }\textfont\ttfam=\eleventt
 \h@big=9.311\p@ \h@Big=12.6\p@ \h@bigg=15.88\p@ \h@Bigg=19.17\p@
 \def\caps{\fam\cpfam \elevencp \f@ntkey=8 }\textfont\cpfam=\elevencp
 \setbox\strutbox=\hbox{\vrule height 9pt depth 4pt width\z@}\relax
 \samef@nt}
\newif\ifelev@n \elev@nfalse
\def\Tenpoint{\tenpoint\twelv@false\elev@nfalse\spaces@t}
\def\Elevenpoint{\elevenpoint\twelv@false\elev@ntrue\spaces@t}
\def\Twelvepoint{\twelvepoint\twelv@true\elev@nfalse\spaces@t}
\def\spaces@t{\relax
\iftwelv@ \ifsingl@\subspaces@t3:4;\else\subspaces@t1:1;\fi
\else \ifelev@n \ifsingl@\subspaces@t2:3;\else\subspaces@t9:10;\fi
\else \ifsingl@\subspaces@t3:5;\else\subspaces@t4:5;\fi \fi \fi
\ifdoubl@ \multiply\baselineskip by 5
\divide\baselineskip by 4 \fi}
\catcode\lq @=12 %
%%%%%%%%%%%%%%%%%%%%%%%%%%%%%%%%%%%%%%%%%%%%%%%%%%%%%%%%%%%%%%%%%%%%%%%%%%%%%
% End of ELEVENPT.TEX
%%%%%%%%%%%%%%%%%%%%%%%%%%%%%%%%%%%%%%%%%%%%%%%%%%%%%%%%%%%%%%%%%%%%%%%%%%%%%
%  This is a macro package REFMACRO.TEX for improved treatment of       %
%   references, footnotes, figures, and tables.                         %
%             By K. Hikasa, March 17, 1987                              %
%%%%%%%%%%%%%%%%%%%%%%%%%%%%%%%%%%%%%%%%%%%%%%%%%%%%%%%%%%%%%%%%%%%%%%%%%
\catcode\lq@=11
\def\keepspacefactor{\let\@sf=\empty \ifhmode
 \edef\@sf{\spacefactor=\the\spacefactor\relax}\relax\fi}
\newcount\footcount \footcount=0
\def\Footnote{\global\advance\footcount by 1 \footnote{\the\footcount}}
\def\footnote#1{\keepspacefactor\refattach{#1}\vfootnote{#1}}

\def\nonfrenchspacing{\sfcode\lq\.=3000 \sfcode\lq\?=3001 \sfcode\lq\!=3001
 \sfcode\lq\:=2000 \sfcode\lq\;=1500 \sfcode\lq\,=1250 }

\nonfrenchspacing
\newmuskip\refskip
\newmuskip\regularrefskip \regularrefskip=2mu
\newmuskip\specialrefskip \specialrefskip=-2mu
\def\refattach#1{\@sf \ifhmode\ifnum\spacefactor=1250 \refskip=\specialrefskip
 \else\ifnum\spacefactor=3000 \refskip=\specialrefskip
 \else\ifnum\spacefactor=1001 \refskip=\specialrefskip
 \else \refskip=\regularrefskip \fi\fi\fi
 \else \refskip=\regularrefskip \fi
 \ref@ttach{\strut^{\mkern\refskip #1}}}
\def\ref@ttach#1{\ifmmode #1\else\ifhmode\unskip${}#1$\relax\fi\fi{\@sf}}
\def\PLrefmark#1{ [#1]{\@sf}}
\def\NPrefmark#1{\refattach{\scriptstyle [ #1 ] }}
\let\PRrefmark=\refattach
\def\refmark{\keepspacefactor\refm@rk}
\def\refm@rk#1{\relax\therefm@rk{#1}}
\def\originalrefs{\let\therefm@rk=\NPrefmark}
\def\PRrefs{\let\therefm@rk=\PRrefmark \let\therefitem=\PRrefitem}
\def\PLrefs{\let\therefm@rk=\PLrefmark \let\therefitem=\PLrefitem}
\def\PRrefitem#1{\refitem{#1.}}
\def\PLrefitem#1{\refitem{[#1]}}
\let\therefitem=\PRrefitem
\def\REF#1{\REFNUM#1%
 \immediate\write\referencewrite{%
 \noexpand\therefitem{#1}}%
\begingroup\obeyendofline\rw@start}
\def\ref{\refnum\?%
 \immediate\write\referencewrite{\noexpand\therefitem{\?}}%
\begingroup\obeyendofline\rw@start}
\def\Ref#1{\refnum#1%
 \immediate\write\referencewrite{\noexpand\therefitem{#1}}%
\begingroup\obeyendofline\rw@start}
\def\REFS#1{\REFNUM#1\global\lastrefsbegincount=\referencecount
\immediate\write\referencewrite{\noexpand\therefitem{#1}}%
\begingroup\obeyendofline\rw@start}
\def\refend{\refm@rk{\number\referencecount}}
{\obeyendofline\gdef\rw@start#1{\def\t@st{#1}\ifx\t@st\blankend%
\endgroup {\@sf} \relax \else \ifx\t@st\bl@nkend \endgroup {\@sf} \relax%
\else \rw@begin#1
\backtotext
\fi \fi } }
\refindent=20pt
\def\REFNUM#1{\eatspace\keepspacefactor\refch@ck \firstreflinetrue%
 \global\advance\referencecount by 1 \xdef#1{\the\referencecount}}
\def\eatspace{\ifhmode\unskip\fi}
\def\refnum#1{\keepspacefactor\refch@ck \firstreflinetrue%
 \global\advance\referencecount by 1 \xdef#1{\the\referencecount}\refend}
\def\figout{\par \penalty-400 \vskip\chapterskip
  \spacecheck\referenceminspace \immediate\closeout\figurewrite
  \figureopenfalse
  \line{\fourteenrm
   \hfil FIGURE CAPTION\ifnum\figurecount=1 \else S \fi\hfil}
  \vskip\headskip
  \input figures.aux
  }
\def\figitem#1{\par\indent \hangindent2\parindent \textindent{Fig. #1\ }}
\def\FIGLABEL#1{\ifnum\number#1<10 \def\figlabel{#1.\zerophant}\else%
\def\figlabel{#1.}\fi}
\def\FIG#1{\FIGNUM#1\FIGLABEL#1%
\immediate\write\figurewrite{\noexpand\figitem{\figlabel}}%
\begingroup\obeyendofline\rw@start}
\def\Figname#1{\FIGNUM#1Fig.~#1\FIGLABEL#1%
\immediate\write\figurewrite{\noexpand\figitem{\figlabel}}%
\begingroup\obeyendofline\rw@start}

\def\fig{\FIGNUM\? fig.~\? \FIGLABEL\?
\immediate\write\figurewrite{\noexpand\figitem{\figlabel}}%
\begingroup\obeyendofline\rw@start}
\def\figure{\FIGNUM\? figure~\? \FIGLABEL\?
\immediate\write\figurewrite{\noexpand\figitem{\figlabel}}%
\begingroup\obeyendofline\rw@start}
\def\Fig{\FIGNUM\? Fig.~\? \FIGLABEL\?
\immediate\write\figurewrite{\noexpand\figitem{\figlabel}}%
\begingroup\obeyendofline\rw@start}
\def\Figure{\FIGNUM\? Figure~\? \FIGLABEL\?
\immediate\write\figurewrite{\noexpand\figitem{\figlabel}}%
\begingroup\obeyendofline\rw@start}
\def\FIGNUM#1{\keepspacefactor\figch@ck \firstreflinetrue%
\global\advance\figurecount by 1 \xdef#1{\the\figurecount}}
\newdimen\digitwidth \setbox0=\hbox{\rm0} \digitwidth=\wd0
\def\zerophant{\kern\digitwidth}
\def\TABNUM#1{\keepspacefactor\tabch@ck \firstreflinetrue%
\global\advance\tablecount by 1 \xdef#1{\the\tablecount}}
\def\tableitem#1{\par\indent \hangindent2\parindent \textindent{Table #1\ }}
\def\TABLE#1{\TABNUM#1\FIGLABEL#1%
\immediate\write\tablewrite{\noexpand\tableitem{\figlabel}}%
\begingroup\obeyendofline\rw@start}
\def\Table{\TABNUM\? Table~\?\FIGLABEL\?%
\immediate\write\tablewrite{\noexpand\tableitem{\figlabel}}%
\begingroup\obeyendofline\rw@start}
\def\tabout{\par \penalty-400 \vskip\chapterskip
  \spacecheck\referenceminspace \immediate\closeout\tablewrite \tableopenfalse
  \line{\fourteenrm\hfil TABLE CAPTION\ifnum\tablecount=1 \else S\fi\hfil}
  \vskip\headskip
  \input tables.aux
  }
\catcode\lq @=12
\PRrefs

%%%%%%%%%%%%%%%%%%%%%%%%%%%%%%%%%%%%%%%%%%%%%%%%%%%%%%%%%%%%%%%%%%%%%%%%%
% End of REFMACRO.TEX
%%%%%%%%%%%%%%%%%%%%%%%%%%%%%%%%%%%%%%%%%%%%%%%%%%%%%%%%%%%%%%%%%%%%%%%%%
%  This is a macro package NEWEQ.TEX for the improved treatment of      %
%   equation numbering.  It replaces EQMACRO.TEX.                       %
%             By K. Hikasa, March 17, 1987  (EQMACRO.TEX)               %
%                      Rev. Jan.  8, 1992                               %
%%%%%%%%%%%%%%%%%%%%%%%%%%%%%%%%%%%%%%%%%%%%%%%%%%%%%%%%%%%%%%%%%%%%%%%%%
\newif\iffinal \finaltrue
\def\showeqname#1{\iffinal\else\hbox to 0pt{\tentt\kern2mm\string#1\hss}\fi}
\def\showEqname#1{\iffinal\else \hskip 0pt plus 1fill
 \hbox to 0pt{\tentt\kern2mm\string#1\hss}\hskip 0pt plus -1fill\fi}
\catcode`\@=11
\def\eqnamedef#1{\relax \ifnum\equanumber<0
\xdef#1{{\noexpand\rm(\number-\equanumber)}}\global\advance\equanumber by -1
    \else \global\advance\equanumber by 1
      \xdef#1{{\noexpand\rm(\chapterlabel \number\equanumber)}}\fi}
\def\eqnamenewdef#1#2{\relax \ifnum\equanumber<0
\xdef#1{{\noexpand\rm(\number-\equanumber#2)}}\global\advance\equanumber by -1
    \else \global\advance\equanumber by 1
      \xdef#1{{\noexpand\rm(\chapterlabel \number\equanumber#2)}}\fi}
\def\eqnameolddef#1#2{\relax \ifnum\equanumber<0
     \global\advance\equanumber by 1
\xdef#1{{\noexpand\rm(\number-\equanumber#2)}}\global\advance\equanumber by -1
    \else \xdef#1{{\noexpand\rm(\chapterlabel \number\equanumber#2)}}\fi}
\def\eqname#1{\eqnamedef{#1}#1}
\def\eqnamenew#1#2{\eqnamenewdef{#1}{#2}#1}
\def\eqnameold#1#2{\eqnameolddef{#1}{#2}#1}
\def\eq{\eqname\lasteq}
\def\eqa{\eqnamenew\lasteq a}
\def\eqb{\eqnameold\lasteq b}
\def\eqc{\eqnameold\lasteq c}
\def\eqd{\eqnameold\lasteq d}
\def\eqnew#1{\eqnamenew\lasteq{#1}}
\def\eqold#1{\eqnameold\lasteq{#1}}
\def\eqn#1{\eqno\eqname{#1}}

%%%%%%%%%%%%%%%%%%%%%%%%%%%%%%%%%%%%%%%%%%%%%%%%%%%%%%%%%%%%%%%%%%
\def\eq@@{\ifinner\let\eqn@=\relax\else\let\eqn@=\eqno\fi\eqn@}
\def\Eq{\eq@@\eq}
\def\Eqnew#1{\eq@@\eqnew{#1}}
\def\Eqold#1{\eq@@\eqold{#1}}
\def\Eqa{\eq@@\eqa}
\def\Eqb{\eq@@\eqb}
\def\Eqc{\eq@@\eqc}
\def\Eqd{\eq@@\eqd}
\def\Eqn#1{\eq@@\eqname{#1}\showeqname{#1}}
\def\Eqnnew#1#2{\eq@@\eqnamenew{#2}{#1}\showeqname{#1}}
\def\Eqnold#1#2{\eq@@\eqnameold{#2}{#1}\showeqname{#1}}
\def\Eqna#1{\eq@@\eqnamenew{#1}a\showeqname{#1}}
\def\Eqnb#1{\eq@@\eqnameold{#1}b\showeqname{#1}}
\def\Eqnc#1{\eq@@\eqnameold{#1}c\showeqname{#1}}
\def\Eqnd#1{\eq@@\eqnameold{#1}d\showeqname{#1}}

\catcode`\@=12
%%%%%%%%%%%%%%%%%%%%%%%%%%%%%%%%%%%%%%%%%%%%%%%%%%%%%%%%%%%%%%%%%%%%%%%%%%
% End of EQMACRO.TEX

\message{ Modified by K. Hikasa: Jan. 9, 1992 Version (PLrefmark modified)}
%%%%%%%%%%%%%%%%%%%%%%%%%%%%%%%%%%%%%%%%%%%%%%%%%%%%%%%%%%%%%%%%%%%%%%%
%%
%
%             Fusion Rules for the $q$-deformed Vertex Operators
%
%                      of $U_q(\widehat{sl(2)})$
%
%                            H. Awata
%
%                            Y. Yamada
%
%                         Written by KEK Phyzzx
%
%%%%%%%%%%%%%%%%%%%%%%%%%%%%  Abstract  %%%%%%%%%%%%%%%%%%%%%%%%%%%%%%%
%%
%     We analyze the properties of the $q$-vertex operators of
%    $U_q(\widehat{sl(2)})$ introduced by Frenkel and Reshetikhin.
%    As the condition for the null vector decoupling,
%    we derive the existence condition of the $q$-vertex operators
%    ( the fusion rules ).
%
%%%%%%%%%%%%%%%%%%%%%%%%   some redefinition  %%%%%%%%%%%%%%%%%%%%%%%%%
%
\def\appendix#1#2{\par\penalty-300\vskip\chapterskip
   \spacecheck\chapterminspace \chapterreset \xdef\chapterlabel{#1}
   \titlestyle{Appendix #2} \nobreak\vskip\headskip \penalty 30000
   \wlog{\string\Appendix\ \chapterlabel} }
\def\Appendix#1{\appendix{#1.}{#1}}
\def\ack{\par\penalty-100\medskip \spacecheck\sectionminspace
   \line{\fourteenbf \hfil  Acknowledgments \hfil}
   \nobreak\vskip\headskip }

\def\pabblock{\begingroup \tabskip=\hsize minus \hsize
   \baselineskip=1.5\ht\strutbox \topspace-2\baselineskip
   \halign to\hsize{\strut ##\hfil\tabskip=0pt\crcr
   \the\Thnum\cr \the\Pubnum\cr
   \the\Month \cr}\endgroup}
%
%%%%%%%%%%%%%%%%%%%%%%%%%%%%%%%%%%%%%%%%%%%%%%%%%%%%%%%%%%%%%%%%%%%%%%%
\overfullrule=0pt
\Thnum={KEK--TH--329}
\Pubnum={KEK Preprint 92--27}
\Month={April 1992}
%%%%%%%%%%%%%%%%%%% title page %%%%%%%%%%%%%%%%%%%%%%%%%%%%%%%%%%%%%%
\pabblock
\vskip1.5cm
\centerline{\fourteenbf
             Fusion Rules for the $q$-Vertex Operators
                      of $U_q(\widehat{sl(2)})$
}
\vskip 2cm
\centerline{ {\twelvecp    Hidetoshi Awata $^1$} {\twelverm and}
             {\twelvecp  \ Yasuhiko Yamada $^2$}  }
           \footnote{1}{E-mail address : awata@jpnkekvm.bitnet}
           \footnote{2}{E-mail address : yamaday@jpnkekvm.bitnet}
\vskip2cm
{\twelvepoint\it
\centerline{$^{1,~2}$
National Laboratory for High Energy Physics {\rm (}KEK{\rm )}}
\centerline{Tsukuba, Ibaraki 305, Japan}
\centerline{and }
\centerline{$^{1}$Dept. of Physics Hokkaido University,}
\centerline{Sapporo 060, Japan}
}
\vskip1.5cm
\centerline {\fourteenbf  Abstract}
\vskip0.5cm
 We analyze the properties of the $q$-vertex operators of
$U_q(\widehat{sl(2)})$ introduced by Frenkel and Reshetikhin.
As the condition for the null vector decoupling,
we derive the existence condition of the $q$-vertex operators
( the fusion rules ).
\endpage

%%%%%%%%%%%%%%%%%%%%%%% ref %%%%%%%%%%%%%%%%%%%%%%%%%%%%%%%%%%%%%%%%%%%

\REF\fr{I.B. Frenkel and N.Yu. Reshetikhin,
       `` Quantum affine algebras and holonomic difference equations.''
		   Preprint (1991).}

\REF\tk{A. Tsuchiya and Y. Kanie,
         Advanced Studies in Pure Math. {\bf 16} (1988) 297,
         Lett. Math. Phys. {\bf 13} (1987) 303.}

\REF\djo{E. Date, M. Jimbo and M. Okado,
       `` Crystal base and $q$-vertex operators ''
	   Osaka Univ. Math. Sci. preprint 1 (1991).}

\REF\dfjmn{B. Davies, O. Foda, M. Jimbo, T. Miwa and A. Nakayashiki,
	`` Diagonalization of the XXZ hamiltonian by vertex operators ''
		RIMS preprint (1992).}

\REF\ay{H. Awata and Y. Yamada,
        `` Fusion rules for the fractional level
		 $\widehat{sl(2)}$ algebra~''
		 KEK Preprint 91-209 KEK-TH-316 (1992),
		  to appear in Mod. Phys. Lett. {\bf A}.}

\REF\s{F.A. Smirnov,
 	   `` Dynamical symmetries of massive integrable models~ ''
	   RIMS preprint 772, 838 (1991).\hfill \break

	   A. LeClair and F.A. Smirnov,
	   `` Infinite quantum group symmetry of fields
	   in massive 2D quantum field theory ''
	   CLNS  preprint 91-1056 (1991).  }

\REF\fl{G. Felder and A. LeClair,
	    `` Restricted quantum affine symmetry
		of perturbed minimal conformal models ''
		RIMS preprint 799 (1991).}

\REF\bs{M. Bauer and N. Sochen,
        `` Fusion and singular vectors
		in $A_1^{(1)}$ highest weight cyclic modules ''
         Saclay preprint (1992).}

\REF\m{F.G. Malikov,
       `` Quantum groups: Singular vectors and BGG resolution ''
	      RIMS Preprint 835 (1991). }

%%%%%%%%%%%%%%%%%%%%%%%%%%%%% 1 %%%%%%%%%%%%%%%%%%%%%%%%%%%%%%%%%%%%%%%
{\bf \chapter{Introduction}}
%%%%%%%%%%%%%%%%%%%%%%%%%%%%%%%%%%%%%%%%%%%%%%%%%%%%%%%%%%%%%%%%%%%%%%%%
%
 The $q$-deformed chiral vertex operators
for the quantum affine Lie algebra $U_q(\widehat{\bf g})$,
defined by Frenkel and Reshetikhin [\fr],
are the $q$-deformed version of the ordinary chiral vertex operators
of the WZNW model introduced by Tsuchiya and Kanie [\tk],
and they have remarkable applications to the solvable lattice models
[\djo, \dfjmn].
 In the present paper, we investigate the fusion rules
for the $q$-vertex operators of $U_q(\widehat{sl(2)})$.

This paper is arranged as follows.
In section 2, we briefly summarize the $U_q(\widehat {sl(2)})$ algebra.
In section 3, we introduce the $q$-vertex operators.
We discuss their matrix elements and $q$-operator product expansion.
Next, in section 4, we present the null vectors explicitly and
calculate the matrix elements of the $q$-vertex operator
including the null vectors.
We give the fusion rules in section 5.
This paper is a $q$-analogue of our previous one [\ay].

%%%%%%%%%%%%%%%%%%%%%%%%%%%%% 2 %%%%%%%%%%%%%%%%%%%%%%%%%%%%%%%%%%%%%%%
{\bf \chapter{ Quantum Algebra $U_q(\widehat {sl(2)})$}}
%%%%%%%%%%%%%%%%%%%%%%%%%%%%%%%%%% 2.1 %%%%%%%%%%%%%%%%%%%%%%%%%%%%%%%%%
\vskip 2mm
\noindent{\bf \S$\,$2.1.}~
 First we define some notation.
The algebra $U_q(\widehat{sl(2)})$ is generated by
$e_i$, $f_i$ and invertible $k_i$ $(i =0,1)$ with relations
$$
\eqalign{
k_i e_j k_i^{-1} = q^{ a_{ij}} e_j &,\hskip 1.3cm
k_i f_j k_i^{-1} = q^{-a_{ij}} f_j  ,\cr
e_i f_j - f_j e_i &= \delta_{ij}{ k_i - k_i^{-1} \over q - q^{-1} },\cr
\sum_{n=0}^{ 1-a_{ij} } (-1)^n
\left[\matrix{ 1-a_{ij} \cr n \cr }\right]
&e_i^{1-a_{ij}-n} e_j e_i^n =0 \hskip1cm (i\neq j) \cr
\sum_{n=0}^{ 1-a_{ij} } (-1)^n
\left[\matrix{ 1-a_{ij} \cr n \cr }\right]
&f_i^{1-a_{ij}-n} f_j f_i^n =0 \hskip1cm (i\neq j) \cr
}\eqn{\algebra}
$$
where
$$
\eqalign{
a_{ij}=\left(\matrix{
\,\, 2 &     -2   \cr
    -2 & \,\, 2   \cr}\right) &, \hskip1.3cm
\left[\matrix{ n \cr m \cr }\right]
={[n]! \over [n-m]! [m]! }, \cr
[n]!=[n] \cdots [1] &, \hskip1.3cm
[n]={q^n-q^{-n} \over q-q^{-1} }.
}\eqn{\qInte}
$$
The algebra $U_q(\widehat{sl(2)})$ is
a Hopf algebra with the comultiplication $\Delta$,
the antipode $S$ and the co-unit $\epsilon$
$$
\eqalign{
\Delta &k_i = k_i \otimes k_i ,\cr
\Delta e_i = e_i \otimes k_i + 1 \otimes e_i &,\hskip 1cm
\Delta f_i = f_i \otimes 1 + k_i^{-1} \otimes f_i,\cr
S(e_i)=-e_i k_i^{-1}, \hskip1cm
S(&k_i)= k_i^{-1}, \hskip1cm
S(f_i)=-k_i f_i,\cr
\epsilon(e_i)=\epsilon(&f_i)=0, \hskip1cm
\epsilon(k_i)= 1.
}\eqn{\Hopf}
$$

 There exists a difference analog of the Virasoro $L_0$ operator
$l_0$ $(=q^{2(k+2)L_0} = p^{L_0} ) \in U_q(\widehat {sl(2)})$
such that $l_0 k_i = k_i l_0$, $l_0 e_i=p^{-\delta_{i,0}}e_i l_0$ and
$l_0 f_i = p^{ \delta_{i,0}} f_i l_0$,
where $p=q^{2(k+2)}$.

%%%%%%%%%%%%%%%%%%%%%%%%%%%%%% 2.2 %%%%%%%%%%%%%%%%%%%%%%%%%%%%%%%%%%%
\vskip 2mm
\noindent{\bf \S$\,$2.2.}~
 Let $M_j$ be the Verma module over $U_q(\widehat{sl(2)})$,
generated by the highest weight vector $\vert j \rangle$, such that
$e_i \vert j \rangle=0$,
$k_1 \vert j \rangle =q^{2j} \vert j \rangle$ and
$k_0 \vert j \rangle =q^{k-2j} \vert j \rangle$.
Here $k \in {\bf C}$ is a level.
Instead of $k$, sometimes we use $t$ $(=k+2)$.
The action of $l_0$ on $M_j$ is defined by
$l_0 \vert j \rangle =p^{h} \vert j \rangle$ with $h_j=j(j+1)/(k+2)$.

 The dual module $M_j^*$ is generated by $ \langle j \vert $
which satisfies
$   \langle j \vert e_i=0 $,
$   \langle j \vert k_1 = q^{2j} \langle j \vert $ and
$   \langle j \vert k_0 = q^{k-2j} \langle j \vert $.
The bilinear form
$ M_j^* \otimes M_j \rightarrow {\bf C} $ is uniquely defined by
$ \langle j \vert j \rangle = 1 $ and
$ \big(  \langle u \vert       X \big) \vert v \rangle =
         \langle u \vert \big( X       \vert v \rangle  \big) $
for any  $\langle u \vert \in M_j^*$, $\vert v \rangle \in M_j$ and
$ X \in U_q(\widehat{sl(2)})$.

 For a generic value of $q$,
the representation theory of $U_q(\widehat{sl(2)})$ is the same as
that of $\widehat{sl(2)}$,
and there are also the null vectors,
which control the reducibility of the representation.
 A null vector $\vert \chi \rangle \in M_j$
(of grade $N$ and charge $Q$ ) is defined by
$e_i \vert \chi \rangle=0$,
$k_1 \vert \chi \rangle=q^{2(j+Q)} \vert \chi \rangle$,
$k_0 \vert \chi \rangle=q^{k-2(j+Q)} \vert \chi \rangle$ and
$l_0 \vert \chi \rangle=p^{h_j+N} \vert \chi \rangle$.
 A null vector $\langle \chi \vert \in M_j^*$
is defined in a similar manner.

%%%%%%%%%%%%%%%%%%%%%%%%%%%%%%%%%%%%%%%%%%%%%%%%%%%%%%%%%%%%%%%%%%%%%

\noindent{\bf  Example.}
Let us give some examples of the null vectors.
Set $2j_{r,s}+1=r-st$, $r,s \in {\bf Z}$ and $t=k+2 \in {\bf C}$.
Examples of the null vectors $\vert \chi_{r,s} \rangle $ are,
$$
\eqalign{
\vert \chi_{r,0} \rangle
&=(f_1)^r \vert j_{r,0} \rangle \qquad {\rm for } \quad r>0 \cr
\vert \chi_{r,0} \rangle
&=(f_0)^{-r} \vert j_{r,0} \rangle \qquad {\rm for } \quad r<0 \cr
\vert \chi_{1,1} \rangle
&=( {     f_0 f_1 f_1 \over [t+1] }
   -{ [2] f_1 f_0 f_1 \over [t  ] }
   +{     f_1 f_1 f_0 \over [t-1] } )\vert j_{1,1} \rangle.
}\eqn\nullex
$$
It is easy to show
$e_1\vert \chi \rangle =0$ and $e_0\vert \chi \rangle =0$.
For example,
$$
e_0 \vert \chi_{1,1} \rangle
=( {[2t+2]\over [t+1]}-{[2][2t]\over [t]}+{[2t-2]\over [t-1]})
f_1f_1\vert j_{1,1} \rangle =0,
\eqno\eq
$$
owing to the following relation ;
${[a]\over [b]}-{[2][a+2]\over [b+1]}+{[a+4]\over[b+2]}
={[2][a-2b]\over [b][b+1][b+2]}$.

%%%%%%%%%%%%%%%%%%%%%%%%%%%%% 3 %%%%%%%%%%%%%%%%%%%%%%%%%%%%%%%%%%%%%%%
{\bf \chapter{ The $q$-deformed Vertex Operator }}
%%%%%%%%%%%%%%%%%%%%%%%%%%%%%%%%%% 3.1 %%%%%%%%%%%%%%%%%%%%%%%%%%%%%%%%%
\vskip 2mm
\noindent{\bf \S$\,$3.1.}~
 For a general Hopf algebra $U$,
the field operator $\phi(z)$ is defined by the following
transformation property under the adjoint action of $U$ :
[\fr, \s, \fl]
$$
{\rm ad} (a) \phi(z) \equiv \sum_k a_k^{(1)} \phi(z) S(a_k^{(2)})
=\rho(a) \phi(z),
\eqn{\trans}
$$
where $a \in U$,
$\Delta^{(n)}(a)=\sum_k a_k^{(1)} \otimes \cdots \otimes a_k^{(n)}$
is the comultiplication $\Delta^{(n)} : U \rightarrow U^{\otimes n}$,
and $\rho$ is a certain representation of $U$.
The adjoint action is the natural generalization of the commutator
for Lie algebras.

 For $U=U_q(\widehat{sl(2)})$,
the $q$-vertex operator $ \phi_{j}(z,x) $
of spin $j$ is defined explicitly as follows
$$
\eqalign{
k_i \phi_j(z,x) &= \rho(k_i) \phi_j(z,x) k_i ,\cr
e_i \phi_j(z,x) &= \rho(e_i) \phi_j(z,x) k_i + \phi_j(z,x) e_i ,\cr
f_i \phi_j(z,x) &= \rho(f_i) \phi_j(z,x)
                 + \rho(k_i^{-1}) \phi_j(z,x) f_i,  \hskip 3cm \, \cr
\noalign{\hbox{and  } }\hskip 3cm \,
l_0 \phi_j(z,x) &= \rho(l_0) \phi_j(z,x) l_0, \cr
}\eqn{\vertex}
$$
where $\rho = \rho_{j,x}$ is a contravariant representation defined by
$$
\eqalign{
\rho(k_1) &=        q^{  2(j - x {d \over dx})}, \hskip2cm
\rho(k_0)  =        q^{ -2(j - x {d \over dx})}, \cr
\rho(e_1) &=          x  [2j - x {d \over dx}] , \hskip1.5cm
\rho(e_0)  = {z \over x} [     x {d \over dx}] , \cr
\rho(f_1) &= {1 \over x} [     x {d \over dx}] , \hskip 2.3cm
\rho(f_0)  = {x \over z} [2j - x {d \over dx}] , \cr
\, \cr
&\, \hskip1.5cm
\rho(l_0) =        p^{         z {d \over dz} }. \cr
}\eqn{\rep}
$$
 Our definition of the $q$-vertex operator is useful
in actual calculations.

In the case of $2j \in {\bf Z}_{\geq 0}$,
$\rho(k_1)$, $\rho(e_1)$ and $\rho(f_1)$
give the $2j+1$ dimensional representation of $U_q(sl(2))$
on $\oplus _{m=0}^{2j} {\bf C} x^m $ .

%%%%%%%%%%%%%%%%%%%%%%%%%%%%%%%%%%%%%%%%%%%%%%%%%%%%%%%%%%%%%%%%%%
\vskip 2mm
\noindent{\bf \S$\,$3.2.}~
 We will investigate the $q$-vertex operator $\phi_{j_2}(z,x)$
in the following two interpretations:
(1) the matrix element and (2) the operator.

\noindent
(1) Matrix element
$ \phi : M_{j_3}^* \otimes M_{j_1} \rightarrow  {\bf C}$.

 From the $ k_1 $ and $ l_0 $ commutation relations,
the ground state matrix element of the $q$-vertex operator is given by
$$
\langle j_3 \vert   \phi_{j_2}(z,x)   \vert j_1 \rangle =
C_{123} z^{ h_{13} } x^{ j_{13} },
\eqn{\matri}
$$
where $h_{13} =-h_1+h_3$, $j_{13} =j_1+j_2-j_3$
and $C_{123}$ is an arbitrary constant.

The other matrix elements for the descendant fields
are uniquely determined by the $U_q(\widehat{sl(2)})$ algebra.
For example,
$$
\eqalign{
\langle j_3 \vert   \phi_{j_2}(z,x)
f_{\alpha_1}\cdots f_{\alpha_n} \vert j_1 \rangle
&= (-1)^n \rho(k_{\alpha_1}) \rho(f_{\alpha_1}) \cdots
\rho(k_{\alpha_n}) \rho(f_{\alpha_n})
\langle j_3 \vert   \phi_{j_2}(z,x) \vert j_1 \rangle,  \cr
\langle j_3 \vert e_{\alpha_1}\cdots e_{\alpha_n}
\phi_{j_2}(z,x)   \vert j_1 \rangle
&= \rho(e_{\alpha_1}) \cdots \rho(e_{\alpha_n})
\langle j_3 \vert \phi_{j_2}(z,x)
k_{\alpha_1}\cdots k_{\alpha_n} \vert j_1 \rangle.  \cr
}\eqno\eq
$$
Note that, if $C_{123}=0$, then $\phi_{j_2}(z,x)=0$.

\endpage
%%%%%%%%%%%%%%%%%%%%%%%%%%%%%%%%%%%%%%%%%%%%%%%%%%%%%%%%%%%%%%%%%%%

\noindent
(2) Operator $ \phi : M_{j_1} \rightarrow  M_{j_3}$.

To study the property of the map
$ \phi : M_{j_1} \rightarrow  M_{j_3}$,
let us consider the image of the ground state
$\vert j_1 \rangle \in M_{j_1}$
( the $q$-operator product expansion ),
$$
\phi_{j_2}(z,x) \vert j_1 \rangle
=\sum_{N,Q}\sum_J \beta_J ^{NQ} q^{ k N - 2j_1 Q }
z^{h_{13}+N} x^{j_{13}-Q} \vert J^{NQ} \rangle_{j_3},
\eqn\OPE
$$
where $\vert J^{NQ} \rangle_{j_3}\in M_{j_3}$
is the basis of the  homogeneous components of
grade $N$ and  charge $Q$.

The coefficients
$\beta_J  =\beta_J^{NQ} $, for fixed $N$, $Q$ in \OPE ,
are determined by the following linear equations,
$$
\sum_J \langle I \vert J \rangle \beta _J
=  \langle I \vert \phi_{j_2}(z,x) \vert j_1\rangle\, \Big\vert_{N,Q},
\eqn\opemat
$$
where $\Big\vert_{N,Q}$ denotes the coefficient of
$q^{ k N - 2j_1 Q } z^{h_{13}+N} x^{j_{13}-Q}$.
In the Appendix,
we give some sample calculations of the coefficients $\beta$.

The vectors
$\vert N,Q \rangle = \sum_J \beta_J^{N,Q} \vert J^{N,Q}\rangle_{j_3}$
are also characterized by the descent equations
$$
\eqalign{
e_1\vert N,Q \rangle &= [-j_1+j_2+j_3+Q+1]\vert N  ,Q+1 \rangle,\cr
e_0\vert N,Q \rangle &= [ j_1+j_2-j_3-Q+1]\vert N-1,Q-1 \rangle.
}\eqno\eq
$$
This is the $q$-analogue of the descent equations in [\bs].

%%%%%%%%%%%%%%%%%%%%%%%%%%%%%%%%%%%%%%%%%%%%%%%%%%%%%%%%%%%%%%%%%
\vskip 2mm
\noindent{\bf \S$\,$3.3.}~
If the determinant of the matrix $\langle I \vert J \rangle  $
( the Shapovalov form ) vanishes,
the consistency of the equation \opemat \
requires the following constraints,
$$
 \langle \chi \vert \phi_{j_2}(z,x) \vert j_1 \rangle =0,
\eqn\nulldecoup
$$
for any null vector $\langle \chi \vert \in M_{j_3}^*$.
This is the so-called the null vector decoupling condition.
Similarly, consistency of the $q$-OPE of
$\phi : M_{j_3}^* \rightarrow M_{j_1}^* $,
i.e. $\langle j_3\vert \phi_{j_2}(z,x) $,
requires $ \langle j_3 \vert \phi_{j_2}(z,x) \vert \chi \rangle=0 $
for all null vectors $\vert \chi \rangle \in M_{j_1}$.
Therefore, the existence of the $q$-vertex operator
depends on the null vector decoupling.

%%%%%%%%%%%%%%%%%%%%%%%%%%%%%%%%%%%%%%%%%%%%%%%%%%%%%%%%%%%%%%%
\noindent
{\bf Example. }
Here we give some examples of the null vector decoupling condition.

\noindent
(i).  For $2j_1+1=r \in {\bf Z}_{>0}$,
by using the null vector $\vert \chi_{r,0} \rangle $ in \nullex ,
$$
\langle j_3 \vert \phi_{j_2} \vert \chi_{r,0} \rangle
=(-1)^r {[j_{13}]! \over [j_{13}-r]!}
q^{(2j_3-2j_1+r+1) r}z^{h_{13}-r}x^{j_{13}}.
\eqno\eq
$$
Hence, the null vector decoupling condition is
$\prod_{n=0}^{r-1} [j_1+j_2-j_3-n]=0$.

\noindent
(ii).  For $2j_1+1=1-t$,
$$
\eqalign{
\langle j_3 \vert \phi_{j_2} \vert \chi_{1,1} \rangle
&=-
( {    [2j_2-j_{13}+2][j_{13}-1][j_{13}] \over [t+1] }
 -{ [2][j_{13}][2j_2-j_{13}+1][j_{13}] \over [t  ] } \cr
&\, \hskip 1cm
 +{    [j_{13}][j_{13}+1][2j_2-j_{13}] \over [t-1] } )
q^{2j_3-2j_1+4}z^{h_{13}-1}x^{j_{13}-1}\cr
=-&{[2][2j_2-j_{13}+t][j_{13}+1-t][2j_2-j_{13}]\over [t+1][t][t-1] }
q^{2j_3-2j_1+4}z^{h_{13}-1}x^{j_{13}-1}.
}\eqno\eq
$$
Here we use the following formula ;
${[a][b]\over [c]}-{[2][a+1][b+1]\over [c+1]}+{[a+2][b+2]\over[c+2]}
={[2][a-c][b-c]\over [c][c+1][c+2]}$.
Therefore, the null vector decoupling condition is
$$
[j_3+j_2-j_1+t][j_1+j_2-j_3+1-t][j_3+j_2-j_1]=0.
\eqn\example
$$

%%%%%%%%%%%%%%%%%%%%%%%%%%%%%%%%%%%% 4 %%%%%%%%%%%%%%%%%%%%%%%%%%%%%%%%
{\bf \chapter{The Null Vectors }}
%%%%%%%%%%%%%%%%%%%%%%%%%%%%%%%%%%%%%%%%%%%%%%%%%%%%%%%%%%%%%%%%%%%%%%%%
%
Here we analyze the null vector decoupling condition
by using the general form of the null vectors
which is given by Malikov [\m].

For \ $t =k+2 \in { \bf C}\setminus \{ 0 \}$
and the highest weight $j$, parametrized as $ 2j_{r,s}+1=r-st $
with $r,s \in {\bf Z}$, such that
(i) $r>0$ and $s \geq 0$ or
(ii) $r<0$ and $s<0$ ,
there exists a unique null vector $\vert \chi_{r,s} \rangle \in M_j$
of grade $N=rs$ and charge $Q=-r$.
And the null vector in $M_{ j_{r,s}} $ is as follows,
$$
\eqalign{
&\vert \chi_{r,s} \rangle =
(f_1)^{r+st} (f_0)^{r+(s-1)t} \cdots \cdots
(f_0)^{r-(s-1)t}(f_1)^{r-st}
\vert j_{r,s} \rangle,
\cr
&\vert \chi_{r,s} \rangle =
(f_0)^{-r-(s+1)t} (f_1)^{-r-(s+2)t} \cdots \cdots
(f_1)^{-r+(s+2)t}(f_0)^{-r+(s+1)t}
\vert j_{r,s} \rangle,
}\eqn\mff
$$
for the cases (i)  and (ii)  respectively.
These formulae make sense after analytic continuation.

%%%%%%%%%%%%%%%%%%%%%%%%%%%%%%%%% 4.2 %%%%%%%%%%%%%%%%%%%%%%%%%%%%%%%%%
%
 We now consider the matrix element including the null vector
$\vert \chi_{r,s} \rangle \in M_{j_1} $.
{}From \vertex , \matri \ and \mff , we obtain
$$
\langle j_3 \vert   \phi_{j_2}(z,x)   \vert \chi_{r,s} \rangle
= (-1)^{(2s+1)r} C_{123} f_{r,s}(j_1,j_2,j_3)
q^{(2j_3-2j_1+r+2s+1)r} z^{ h_{13}-rs } x^{ j_{13}-r },
\eqno\eq
$$
where
$$
\eqalign{
&f_{r,s}(j_1,j_2,j_3)=
\prod_{n=0}^{r-1} \prod_{m=0}^{s} [ j_1 + j_2 - j_3 - n + mt ]
\prod_{n=1}^{r}   \prod_{m=1}^{s} [-j_1 + j_2 + j_3 + n - mt ],
\cr
&f_{r,s}(j_1,j_2,j_3)=
\prod_{n=0}^{-r-1} \prod_{m=0}^{-s-1} [-j_1 + j_2 + j_3 - n + mt ]
\prod_{n=1}^{-r}   \prod_{m=1}^{-s-1} [ j_1 + j_2 - j_3 + n - mt ].
}\eqn\nullmatrix
$$
for the cases (i) and (ii)  respectively.
The proof is given by a method similar to that used in [\ay].
Here we use, for instance,
$$
({ 1 \over x } [ x \partial ] )^n x^j
= { [j]! \over [j-n]! } x^{j-n},
$$
for any $n,j \in {\bf C}$.
The formulae \nullmatrix \ are our main results.
Note that similar expressions are obtained
for a projection of the singular vectors [\m] .

%%%%%%%%%%%%%%%%%%%%%%%%%%%%%%%%% 4.3 %%%%%%%%%%%%%%%%%%%%%%%%%%%%%%%%%
%
 Next, we consider another matrix element
$\langle \chi_{r,s} \vert   \phi_{j_2}(z,x)   \vert j_1 \rangle $.
This element follows from
$\langle j_3 \vert   \phi_{j_2}(z,x)   \vert \chi_{r,s} \rangle $,
because of the existence of the following anti-algebra automorphism.
Indeed, the $\sigma$ defined by
$$
\eqalign{
&\sigma (e_i)=-q^{-2} f_i k_i, \quad
\sigma (k_i)= k_i,  \quad
\sigma (f_i)=-q^2 k_i^{-1} e_i,  \cr
\sigma \big(  dx^{-j} \phi_j(z,x) \big) &=
 dy^{-j} \phi_j(w,y),  \quad
\sigma \big(
dx^{-j} \langle u \vert   \phi_{j}(z,x)   \vert v \rangle \big)
= dx^{-j} \langle u \vert   \phi_{j}(z,x)   \vert v \rangle,
}\eqno\eq
$$
with $ w = z^{-1} $ and $ y = x^{-1} $,
is the anti-algebra automorphism, which means \break
$\sigma(k_i^{-1}) \,  \sigma(e_j) \, \sigma(k_i) \,
= \,  q^{ a_{ij}} \, \sigma(e_j)$ ,
$\Delta \, \sigma(e_i) =
\sigma(e_i) \otimes \sigma(k_i) + \sigma(1) \otimes \sigma(e_i)$
and \break
$\sigma(\phi_j(z,x))\sigma(k_i) =
\sigma(k_i) \rho(k_i) \sigma(\phi_j(z,x))$ etc.
Consequently, we obtain
$$
\langle \chi_{r,s} \vert   \phi_{j_2}(z,x)   \vert j_1 \rangle
= (-1)^{(2s+1)r} C_{123} f_{r,s}(j_3,j_2,j_1)
q^{(2j_1 -2j_3+r+2s+1)r} z^{ h_{13}+rs } x^{ j_{13}+r },
\eqno\eq
$$
with the same function $f_{r,s}(j_3,j_2,j_1)$ as \nullmatrix.

%%%%%%%%%%%%%%%%%%%%%%%%%%%%% 5 %%%%%%%%%%%%%%%%%%%%%%%%%%%%%%%%%%%%%%%
{\bf \chapter{ Fusion Rules }}
%%%%%%%%%%%%%%%%%%%%%%%%%%%%%%%%%%%%%%%%%%%%%%%%%%%%%%%%%%%%%%%%%%%%%%%
%
 The factor $f_{rs}(j_1,j_2,j_3)$ is written
in termes of the $q$-integer;
however, this factor is almost the same as
that of the $\widehat{sl(2)}$ case [\ay].
Therefore, the fusion rules are similar to the $\widehat{sl(2)}$ case,
if $q$ is not a root of unity.

\vskip 2mm
\noindent{\bf \S$\,$5.1.}~
When $M_{j_1}$ has a null vector $\vert \chi_{r,s} \rangle$,
there exists a $q$-vertex operator
if and only if the matrix element including the null vector vanishes,
$\langle j_3 \vert   \phi_{j_2}(z,x)   \vert \chi_{r,s} \rangle = 0$,
i.e. $f_{r,s}(j_1,j_2,j_3) =0$.
And this gives the fusion rules.

In the case that $M_{j_3}^*$ also has a null vector
$\langle \chi_{r_3,s_3} \vert$,
the existence condition of the  $q$-vertex operator is
$\langle j_3 \vert   \phi_{j_2}(z,x)   \vert \chi_{r_1,s_1}\rangle = $
$\langle \chi_{r_3,s_3} \vert  \phi_{j_2}(z,x) \vert j_1 \rangle = 0$,
i.e. $f_{r_1,s_1}(j_1,j_2,j_3)=f_{r_3,s_3}(j_3,j_2,j_1)=0$.

%%%%%%%%%%%%%%%%%%%%%%%%%% 5.2 %%%%%%%%%%%%%%%%%%%%%%%%%%%%%%%%%%%%
\vskip 2mm
\noindent{\bf \S$\,$5.2.}~
 If the level is rational $t=p/q$,
with the coprime integers $p$ and $q$, then  $j_{r,s} = j_{r-p,s-q}$.
Hence, there are two independent null vectors
for  $ j_1 = j_{r,s} = j_{r-p,s-q} $.
There exists a $q$-vertex operator if and only if
$\langle j_3 \vert   \phi_{j_2}(z,x)   \vert \chi_{r,s} \rangle =
\langle j_3 \vert   \phi_{j_2}(z,x)   \vert \chi_{r-p,s-q} \rangle=0$,
i.e. $f_{r,s}(j_1,j_2,j_3)=f_{r-p,s-q}(j_1,j_2,j_3)=0$.

The last fusion rule says that $M_{j_3}$ also has two null vectors.
Hence, we also need the condition
$\langle \chi_{r,s} \vert   \phi_{j_2}(z,x)   \vert j_1 \rangle =
\langle \chi_{r-p,s-q} \vert   \phi_{j_2}(z,x) \vert j_1 \rangle = 0$,
that is \break
$f_{r,s}(j_3,j_2,j_1)$  $=f_{r-p,s-q}(j_3,j_2,j_1)=0$.
However, this condition is included in the above one.

In Ref [\djo], it is shown that the space of vertex operators has
a basis indexed by admissible triples
$\{ \Phi_{\lambda}^{\mu b}(z) \}$,
for dominant integral weights $\lambda$ and $\mu$
for any $U_q(\widehat {\bf g})$.
For the integrable case, i.e. $k$ is a non-negative integer,
our results agree with those of Ref [\djo].

%%%%%%%%%%%%%%%%%%%%%%%%%%%%%%%%%%%%%%%%%%%%%%%%%%%%%%%%%%%%%%%%%%%%%%
{\bf \chapter{Conclusions and Remarks}}
%%%%%%%%%%%%%%%%%%%%%%%%%%%%%%%%%%%%%%%%%%%%%%%%%%%%%%%%%%%%%%%%%%%%%%

 As the condition for the null vector decoupling,
we have derived the fusion rules
for the $U_q(\widehat{sl(2)})$ algebra.
Similar results are expected for other quantum affine Lie algebras.
When $q$ is a root of unity,
the structure of the representations changes drastically,
and we need further investigation to understand
the $q$-vertex operators in such cases.

 %%%%%%%%%%%%%%%%%%%%%%%%%%%%%%%%%%%%%%%%%%%%%%%%%%%%%%%%%%%%%%%%%%%
{\bf \ack}
We would like to thank F. G. Malikov, A. Tsuchiya
and the members of KEK theory group for valuable discussions.
We would also like to thank to N. A. McDougall
for a careful reading of the manuscript.

%%%%%%%%%%%%%%%%%%%%%%%%%%%%%%%%%%%%%%%%%%%%%%%%%%%%%%%%%%%%%%%%%%%%%%%
{\bf \appendix{A.}{} }
%%%%%%%%%%%%%%%%%%%%%%%%%%%%%%%%%%%%%%%%%%%%%%%%%%%%%%%%%%%%%%%%%%%%%%%%

Let us give some examples of the $q$-OPE calculation.
Set $a\equiv k-2j_3$, $b\equiv 2j_3$ and $j_{13} \equiv j_1+j_2-j_3$.
In \OPE  , let
$$
\sum_J \beta_J ^{NQ} \vert J^{NQ} \rangle_{j_3}
=\sum_{\alpha_1,\cdots,\alpha_n} \beta_{\alpha_1,\cdots,\alpha_n}
f_{\alpha_1} \cdots f_{\alpha_n}   \vert j_3 \rangle
\eqno\eq
$$
where
$N=\sum_i \bar\alpha_i$ , $Q=\sum_i \bar\alpha_i -\sum_i \alpha_i$ ,
$\alpha_i=0,1$ and $\bar \alpha=1-\alpha$.
\footnote\dag{In general, such a basis is an over complete one.}

\noindent
(i).  For $N=0$ and $Q=-r \in {\bf Z}$, the basis is
$(f_1)^r \vert j_3 \rangle $,
the Shapovalov form is
$\langle j_3 \vert (e_1)^r (f_1)^r \vert j_3 \rangle
=  [r]! [b]! / [b-r]!  $
and
$\langle j_3 \vert (e_1)^r \phi_{j_2}(z,x) \vert j_1 \rangle$
$=  q^{2j_1 r} z^{h_{13}} x^{j_{13}+r} $
$  [2j_2-j_{13}]! / [2j_2-j_{13}-r]! $.
Hence, we obtain
$$
{  [r]! [b]! \over [b-r]!  }
\beta_{\underbrace{1,\cdots,1}_{r \rm \; times}}=
{  [2j_2-j_{13}]! \over [2j_2-j_{13}-r]! }.
\eqno\eq
$$
If $[b-r+1]=[2j_3-r+1]=0$,
then $\langle j_3 \vert (e_1)^r $ is a null vector;
therefore the existence condition of the $q$-vertex operator
$ \phi_{j_2} : M_{j_1} \rightarrow  M_{j_3}$ is
$ \prod_{n=0}^{r-1} [j_3+j_2-j_1 -n] =0$.

\noindent
(ii). For $N=1$ and $Q=0$, the basis is
$f_1 f_0 \vert j_3 \rangle $ and $f_0 f_1 \vert j_3 \rangle $
and the Shapovalov form is
$$
\left( \matrix{
 \langle j_3 \vert e_1 e_0 f_1 f_0 \vert j_3 \rangle
&\langle j_3 \vert e_1 e_0 f_0 f_1 \vert j_3 \rangle \cr
 \langle j_3 \vert e_0 e_1 f_1 f_0 \vert j_3 \rangle
&\langle j_3 \vert e_0 e_1 f_0 f_1 \vert j_3 \rangle \cr }\right)
= \left( \matrix{
[a][b] &[a+2][b] \cr
[a][b+2] &[a][b] \cr }\right).
\eqno\eq
$$
{}From \opemat , we can obtain the $\beta$
$$
[2][a][b][a+b+2]
\left( \matrix{\beta_{10} \cr \beta_{01} \cr }\right) =
\left( \matrix{ -[a][b]    &[a+2][b] \cr
                [a][b+2]   &-[a][b]   \cr }\right)
\left( \matrix{ [j_{13}+1][2j_2-j_{13}]\cr
                [2j_2-j_{13}+1][j_{13}]\cr } \right).
\eqno\eq
$$

\noindent
(iii).  For $N=1$ and $Q=-1$, the basis is
$f_1 f_1 f_0 \vert j_3 \rangle $, $f_1 f_0 f_1 \vert j_3 \rangle $
and $f_0 f_1 f_1 \vert j_3 \rangle $,
and the Shapovalov form is
$$
[2]\left( \matrix{
[a][b-1][b  ] &[a+2][b-1][b  ] &[a+4][b-1][b] \cr
[a][b  ][b+1] &[a+1][b  ][b  ] &[a+2][b-1][b] \cr
[a][b+1][b+2] &[a  ][b  ][b+1] &[a  ][b-1][b] \cr }\right).
\eqno\eq
$$
Therefore, we have
$$
\eqalign{
&[2]^2[a][b][b-1][a+b+2][a+2b+2]
\left( \matrix{\beta_{110} \cr \beta_{101} \cr \beta_{011} \cr }\right)
= \cr
&\, \cr
&\left( \matrix{
  \matrix{[a][b][b-1]\cr
  \times [a+b+2] \cr}
& \matrix{-[2][a][b][b-1]\cr
  \times [a+b+3] \cr }
& \matrix{[b][b-1]\cr
  \times \left( \matrix {[2][a+2][a+b+3]\cr -[a+4][a+b+2]\cr}\right)}
\cr
\ & \ & \ \cr
\ & \ & \ \cr
  \matrix {-[2][a][b-1][b+1]\cr
  \times [a+b+2] \cr}
& \matrix{[2][a][b-1]\cr
  \times \left(\matrix {[b-1][a+b+2]\cr +[b+2][a+b+3]\cr}\right)\cr}
& \matrix{-[2][a][b][b-1]\cr
  \times [a+b+3] \cr } \cr
\ & \ & \ \cr
\ & \ & \ \cr
  \matrix{[a][b][b+1]\cr
  \times [a+b+2] \cr}
& \matrix{-[2][a][b-1][b+1]\cr
  \times [a+b+2]\cr }
& \matrix{[a][b][b-1]\cr
  \times [a+b+2] \cr}\cr} \right)\cr
}$$
$$
\times
\left( \matrix{ [j_{13}+2]   [2j_2-j_{13}-1][2j_2-j_{13}] \cr
                [2j_2-j_{13}][j_{13}+1]     [2j_2-j_{13}] \cr
                [2j_2-j_{13}][2j_2-j_{13}+1][j_{13}] \cr } \right).
\eqno\eq
$$

If the determinant of the Shapovalov form vanishes, i.e.
$[a+2b+2]=[k+2j_3+2]=0$,
then there is a null vector $\langle \chi \vert $ such as
$$
\langle \chi \vert = \langle j_3 \vert
 (  {     e_1 e_1 e_0 \over [t+1] }
   -{ [2] e_1 e_0 e_1 \over [t  ] }
   +{     e_0 e_1 e_1 \over [t-1] } ).
\eqno\eq
$$
Hence the existence condition of the $q$-vertex operator is
$[j_1+j_2- j_3 + t] [j_3 + j_2 - j_1 + 1 - t] [j_1+j_2-j_3] = 0$ .

\noindent
(iv). Finally, there is a following relation
$$
\beta_{\alpha_1,\cdots,\alpha_n}(j_1,j_2,j_3)=
\beta_{\bar\alpha_1,\cdots,\bar\alpha_n}
({k\over 2}-j_1,j_2,{k\over 2}-j_3).
\eqno\eq
$$

\par \penalty-400 \vskip\chapterskip
   \spacecheck\referenceminspace \immediate\closeout\referencewrite
   \referenceopenfalse
   \line{\fourteenbf \hfil  References \hfil}\vskip\headskip
   \input reference.aux   

\bye